\documentstyle[12pt,aps,preprint,epsf]{revtex}
\begin{document}
%%%%%%%%%%%%%%%%%%%%%%%%%%%%%%%%%%%%%%%%%%%%%%%%%%%%%%%%%%%%%%%%
%
%
%
%%%%%%%%%%%%%%%%%%%%%%%%%%%%%%%%%%%%%%%%%%%%%%%%%%%%%%%%%%%%%%%%
%%%%%%%%%%%%%%%%%%%%%%%DEFINITION%%%%%%%%%%%%%%%%%%%%%%%%%%%%%%%
%%%%%%%%%%%%%%%%%%%%%%%%%%%%%%%%%%%%%%%%%%%%%%%%%%%%%%%%%%%%%%%%
\def\figurename{Fig.}
\def\X{\mbox{\boldmath$X$}}
\def\Y{\mbox{\boldmath$Y$}}
\def\P{\mbox{\boldmath$P$}}
\def\x{\varphi_1}
\def\y{\varphi_2}
\def\vpi{\varphi_i}
\def\vpj{\varphi_j}
\def\vx{\varphi}
\def\e{\epsilon}
%%%%%%%%%%%%%%%%%%%%%%%%%%%%%%%%%%%%%%%%%%%%%%%%%%%%%%%%%%%%%%%%
%
%
%
%%%%%%%%%%%%%%%%%%%%%%%%%%%%%%%%%%%%%%%%%%%%%%%%%%%%%%%%%%%%%%%%
%
%
%
\rightline{\large\baselineskip20pt\rm\vbox to10pt{
\baselineskip14pt
\hbox{DPNU-97-37}
\hbox{Aug. 1997}
}}
\vspace{0.5cm} 
\begin{center}
{\Large\bf Cosmological perturbation with two scalar fields}\\
{\Large\bf in reheating after inflation} 
\end{center}
\centerline{\large
A.Taruya\footnote{e-mail:~ataruya@allegro.phys.nagoya-u.ac.jp}
 and Y.Nambu \footnote{e-mail:~nambu@allegro.phys.nagoya-u.ac.jp}
 }
\begin{center}
{\em Department of Physics, Nagoya University, Chikusa-ku, 
Nagoya 464-01, Japan }
\end{center}
\bigskip
%
%
%
%%%%%%%%%%%%%%%%%%%%%%%%%%%%%%%%%%%%%%%%%%%%%%%%%%%%%%%%%%%%%%%%
\begin{center}
{\large\bf Abstract}
\end{center}

We investigate the cosmological perturbation of two-scalar field model 
during the reheating phase after inflation. Using the exact solution of the 
perturbation in long-wavelength limit, which is expressed in terms of the 
background quantities, we analyze the behavior of the metric 
perturbation. The oscillating inflaton field gives rise to the 
parametric resonance of the massless scalar field and this leads to 
the amplification of the iso-curvature mode of the metric perturbations.
%%%%%%%%%%%%%%%%%%%%%%%%%%%%%%%%%%%%%%%%%%%%%%%%%%%%%%%%%%%%%%%%
%
%
\\
\vspace{0.1cm}

\newpage
%
%
%%%%%%%%%%%%%%%%%%%%%%%%%%%%%%%%%%%%%%%%%%%%%%%%%%%%%%%%%%%%%%%%
\section{Introduction}
%%%%%%%%%%%%%%%%%%%%%%%%%%%%%%%%%%%%%%%%%%%%%%%%%%%%%%%%%%%%%%%%
%
%
Recently, a theory of reheating after inflation is developed and 
importance of the oscillating scalar fields is recognized
\cite{Dolgov,Traschen,KFL1,STB}. 
The non-linear interaction between the scalar fields 
amplifies the fluctuations of the scalar fields by the effect of 
parametric resonance and the energy of the 
oscillating field is transfered to the fluctuation of the massless field by 
catastrophic particle production. 
A large amount of studies about the reheating process 
has been done\cite{KFL2}. 
From the viewpoint of the structure formation of our universe, one of the 
most important question is whether the growth of 
fluctuations of the matter field during reheating affects the large-scale 
inhomogeneities in the universe or not. To answer this question, 
we must study the evolution of the metric and the matter fluctuation 
in general relativistic treatment. 

Concerning the works on the theory of cosmological perturbation, 
several authors studied a single scalar field model 
in the context of the reheating scenario
and investigated a role of the coherent oscillating 
field on the evolution of large-scale structure in the universe
\cite{ynambu,KandH,hwang1}.
Hamazaki and Kodama analyze the metric 
perturbation in the model with two-component fluid 
and discuss the effect of parametric resonance\cite{HandK}. 
They evaluate the curvature perturbation $\zeta$ (Bardeen parameter) 
by replacing the scalar fields 
with the perfect fluids. 
The effect of parametric resonance is all included in the energy 
transfer term which 
should be given by phenomenological description. 
They conclude that $\zeta$ is well-conserved during 
reheating and the parametric resonant decay 
does not affect the scenario of the structure formation. 
Although their conclusion seems to be valid in the old version of the 
reheating scenario, which is dominated by the Born decay process, 
it is not clear that their analysis is appropriate 
when the non-linear evolution of the scalar fields may play an important role. 
We should keep in mind that the parametric resonance does not occur in 
the perfect fluid system.  

In this letter, 
we analyze the cosmological perturbation of the scalar field model 
to clarify whether the dynamics of coherently oscillating scalar 
field affects the metric perturbation or not. 
The model we treat here contains two scalar fields $\phi, \chi$ 
with the potential $V(\phi,\chi)=m^2\phi^2/2+g^2\phi^2\chi^2/2$  
where $\phi$ is the inflaton and $\chi$ is the massless boson field. 
%
%
%
%
%%%%%%%%%%%%%%%%%%%%%%%%%%%%%%%%%%%%%%%%%%%%%%%%%%%%%%%%%%%%%%%%
\section{Basic equations}
%%%%%%%%%%%%%%%%%%%%%%%%%%%%%%%%%%%%%%%%%%%%%%%%%%%%%%%%%%%%%%%%
%
%
We consider a homogeneous, isotropic and flat cosmological model with 
scalar type metric perturbations. In longitudinal (Newtonian) gauge, the 
metric is given by 
\begin{equation}
ds^2=-(1+2\Phi)dt^2+a^2(t)(1-2\Psi)\delta_{ij}dx^idx^j.
  \label{metric}
\end{equation}
$\Phi, \Psi$ correspond to the gauge-invariant potential, which 
characterize the physical mode of the metric perturbations. 
The Bardeen parameter which 
represents the spatial curvature perturbation on the uniform energy 
density slice in super-horizon scale is written by
\cite{Lyth,Salopek,MFB,BandW}  
\begin{equation}
  \zeta=\Phi-\frac{H^2}{\dot{H}}(\Phi+H^{-1}\dot{\Phi})~;
  ~~H\equiv\frac{\dot{a}}{a}.
  \label{bardeen1}
\end{equation}
%
%%%%%%%
For the later analyses, we use the following dimensionless variables:
\begin{equation}
  \tau=mt,~~
  h=\frac{H}{m},~~
  \x=\sqrt{\frac{4\pi}{3}}\frac{\phi}{m_{pl}},~~
  \y=\sqrt{\frac{4\pi}{3}}\frac{\chi}{m_{pl}},~~
  \lambda=\frac{3}{4\pi}\left(\frac{gm_{pl}}{m}\right)^2. 
\end{equation}
%%%%%%%
In terms of these variables, the background Einstein equations are given by 
\begin{eqnarray}
  &&h^2=\x'^2+\y'^2+2 U(\x,\y);~~
  U=\frac{1}{2}\x^2+\frac{1}{2}\lambda \x^2\y^2, 
  \label{bg-eq1}
\\
&&\vpi''+3h\vpi'+\frac{\partial U}{\partial \vpi}=0~~~(i=1,2),
\label{bg-eq2}
\end{eqnarray}
where $(')=d/d\tau$.
%%%%%%%
We use the following gauge-invariant variable to describe the 
perturbation:
\begin{equation}
  Q_i=\delta \vpi + \frac{\vpi'}{h} \Psi,  
  \label{Mukhanov-Q}
\end{equation}
where  $\delta \vpi$ is perturbation of the scalar field $\vpi$. 
The variable $Q_i$ was first introduced by Mukhanov 
\cite{Mukhanov} and all the other gauge-invariant quantities 
can be expressed by these variables. Using the fact $\Phi=\Psi$ 
in the present model, the Bardeen parameter defined by 
(\ref{bardeen1}) is written by 
\begin{equation}
  \zeta= \frac{h}{\x'^2+\y'^2}(\x'Q_1+\y'Q_2).
\label{bardeen2}
\end{equation}
%
%%%%%%%
The evolution equation for $Q_i$ is simply expressed by 
\cite{hwang2}
\begin{equation}
Q_i''+3hQ_i'+\left(\frac{k}{ma}\right)^2 Q_i
+\sum_{j=1}^2
\left[\frac{\partial^2 U}{\partial \vpi\partial \vpj}
-\frac{6}{a^3}\left(\frac{a^3}{h}\vpi'\vpj'\right)'~
\right]Q_j=0.
  \label{Mukhanov-eqs}
\end{equation}
This is our basic equation.
%
%
%%%%%%%%%%%%%%%%%%%%%%%%%%%%%%%%%%%%%%%%%%%%%%%%%%%%%%%%%%%%%%%%
\section{Solution in long wavelength limit}
\label{sec: k=0-sol}
%%%%%%%%%%%%%%%%%%%%%%%%%%%%%%%%%%%%%%%%%%%%%%%%%%%%%%%%%%%%%%%%
%
%
%%%%%%%
As we are interested in the formation of the 
large-scale structure in our universe, 
it is important to investigate the metric fluctuation 
outside the scale of the Hubble horizon. Fortunately, we can obtain 
the exact solution of eq.(\ref{Mukhanov-eqs}) in the long wavelength limit 
($k\to0$) and the result is expressed in terms of the background quantities. 

Let us suppose the solution of the background field is written as 
$\vpi(\alpha,~C)$,  
where $\alpha\equiv\log{a}$ and $C$ is the parameter given by a 
suitable choice of the integration constants. 
$\alpha$ plays a role of the time parameter and 
we can trace the time evolution of the trajectory of the background 
solution in configuration space $(\x,\y)$ using this parameter. 
$C$ distinguishes trajectories in this space 
and remains constant on a each trajectory. 
%%%%%%%%
%Rewriting the field equation (\ref{bg-eq2}) by using $\alpha$ and 
%differentiating with respect to the parameter $\alpha$, $C$,  
%we can show that the tangential vectors 
%$d\vpi/d\alpha$, $d\vpi/dC$ satisfy (\ref{Mukhanov-eqs}) 
%in the limit $k\to0$. 
%Therefore these vectors are the independent solutions of 
%(\ref{Mukhanov-eqs}) in long wavelength limit\cite{ataruya}. 
%%%%%%%
Using the time parameter $\alpha$, the field equation (\ref{bg-eq2}) 
is rewritten as
\begin{equation}
  \frac{d^2\vpi}{d\alpha^2}+\left(3+h^{-1}\frac{dh}{d\alpha}\right)
\frac{d\vpi}{d\alpha}+h^{-2}\frac{\partial U}{\partial\vpi}=0.
\label{field-eq-alpha}
\end{equation}
Differentiating (\ref{field-eq-alpha}) with respect to $\alpha, C$, 
we can show that the equation of motion for the tangent vectors 
$d\vpi/d\alpha$, $d\vpi/dC$ becomes (\ref{Mukhanov-eqs}) in the limit 
$k\to0$.
Therefore these vectors are the independent solutions of 
(\ref{Mukhanov-eqs}) in long wavelength limit\cite{ataruya}. 
%%%%%%%

Since (\ref{Mukhanov-eqs}) are the second order coupled differential 
equations, we have four independent solutions. The remaining two 
solutions can be obtained as follows:
define the component of a $2\times2$ matrix $\X$ as 

\begin{equation}
  \X_{11}=\frac{d\x}{d\alpha},~~  \X_{12}=\frac{d\x}{dC},~~
  \X_{21}=\frac{d\y}{d\alpha},~~  \X_{22}=\frac{d\y}{dC}. 
\end{equation}
Each row vectors in $\X$ satisfy (\ref{Mukhanov-eqs}). 
Expressing the remaining solutions by the matrix $\Y$ and putting 
$\Y=\X\cdot\P$ yields
\begin{equation}
  \P''+(3h+2\X^{-1}\X')\P'=\mbox{\boldmath$O$}.
\end{equation}
We can obtain $\P'=\X^{-1}(\X^{-1})^T/a^3$ by using the fact that 
$\X'\cdot\X^{-1}$ is the symmetric matrix\cite{ataruya}. 
Thus the general solution of (\ref{Mukhanov-eqs}) in long 
wavelength limit is given by 
\begin{equation}
  Q_i=\sum_{j=1}^2
\left(
c_j^+ \X_{ij}+c_j^-\left[\X\int \frac{d\tau}{a^3} 
\X^{-1}(\X^{-1})^T\right]_{ij}
\right), 
\label{k=0-sol}
\end{equation}
where $(^T)$ represents transpose of the matrix and $c_j^{\pm} (j=1,2)$ are 
arbitrary constants. 
%
%
%
%
%%%%%%%%%%%%%%%%%%%%%%%%%%%%%%%%%%%%%%%%%%%%%%%%%%%%%%%%%%%%%%%%
\section{Background dynamics in reheating phase}
\label{sec: bg-evolution}
%%%%%%%%%%%%%%%%%%%%%%%%%%%%%%%%%%%%%%%%%%%%%%%%%%%%%%%%%%%%%%%%
%
%
%
We can know the behavior of the long wavelength perturbation from 
(\ref{k=0-sol}) if we obtainthe background solution.
%%%%%%% 
As we pay our attention to the reheating phase of 
the inflaton dominated universe, we assume the condition 
$\x,\y\lesssim1$ and $\lambda \y^2\lesssim1$. The former is the condition 
for the oscillation of the scalar fields. The latter condition implies 
that the potential energy is dominated by the massive inflaton. 

%%%%%%% 
We first solve the background equations in a naive treatment.   
Because $\lambda \y^2\lesssim1$, the evolution equations 
   (\ref{bg-eq1})(\ref{bg-eq2}) can be solved separately and we get 
   oscillatory behavior of the inflaton 
   $\x\simeq \bar{\varphi}(\tau)\cos{\tau}$, 
   where $\bar{\varphi}\propto 1/\tau$. 
   Substituting this into (\ref{bg-eq2}) again, we find that the 
   equation for the field $\y$ becomes the same form as the Mathieu equation: 
\begin{equation}
(a^{3/2}\y)''+[A+2q\cos{(2\tau)}](a^{3/2}\y)=0~;~a\propto\tau^{2/3}, 
  \label{mathieu}
\end{equation}
    where the coefficients $A$ and $q$ 
   are time dependent functions given by 
   $A=2q=\lambda \bar{\varphi}^2(\tau)/2 $. 
   The stability/instability chart of the Mathieu equation says that $\y$ 
   has unstable behavior at $1/3\leq q\leq 1$, 
   in the region of the first resonance band of the Mathieu function
   \cite{ataruya}. 

%%%%%%% 
To investigate the metric perturbation during the reheating, we must 
  study the background evolution more precisely. We shall analyze the 
  background dynamics by using the {\it renormalization group} 
  (RG) {\it method}. 
%%%%%%% 
RG method is a technique of asymptotic analysis which improves the result 
of the naive perturbation and it provides an unified approach to solve 
differential equations including singular perturbation method 
\cite{Oono}\cite{kunihiro}. 

%%%%%%% 
To apply the RG method, we identify the small expansion parameter $\e$ 
as the amplitude of the scalar fields $\vpi\sim{\cal O}(\e)<1$. 
Since we want to take into account of the non-linear interaction, 
we assume $\tilde{\lambda}\equiv \e \lambda\sim{\cal O}(1)$ which 
implies $\lambda$ is not small parameter. Then the background quantities 
are expanded as follows:
\begin{equation}
  \vpi=\e \vx_{i(1)}+\e^2 \vx_{i(2)}+ \cdots,~~~
    h=\e h_{(1)}+\e^2 h_{(2)}+ \cdots. 
    \label{e-expansion}
  \end{equation}
%
%%%%%%% 
The crucial treatment in the present model is to insert the mass term 
$\omega^2\y$ in the equation of $\y$ (\ref{bg-eq2}). 
We put $\omega^2=1+\e \sigma$ and 
take the limit $\e\sigma\to-1$ after the end of calculation. This treatment is 
necessary to extract the effect of parametric resonance in $\y$ field. 

%%%%%%% 
In appendix, we apply the RG method to the 
background equations (\ref{bg-eq1})(\ref{bg-eq2}).
The final form of the solution up to ${\cal O}(\e)$ becomes
\begin{equation}
\x\simeq h\cos{\Theta}\cos{(\tau+\theta)},~~
\y\simeq h\sin{\Theta}\cos{(\tau+\psi)}, 
  \label{RG-sol}
\end{equation}
where we replace $\e h_{(1)}$ with $h$. The variables $\Theta,\theta,\psi$ 
and $h$ are time dependent and determined by so-called RG equations 
(\ref{RG-1})-(\ref{RG-4}) which are the result of the renormalization 
of the initial amplitude. 
%
%

%%%%%%% 
We further impose the condition $\y \ll \x$ ($\Theta\ll1$),   
which makes our analysis easier. 
We will restrict our analysis in the case of $\Theta\ll1$ hereafter.  
(\ref{RG-3}) becomes $\theta'\simeq 0$ 
and we can set $\theta=0$ without 
loss of generality. Eqs.(\ref{RG-1})(\ref{RG-2})(\ref{RG-4}) become
\begin{eqnarray}
&&\Theta'\simeq \frac{1}{2}q(\tau)\sin{\gamma}\cdot\Theta,~~~
\gamma'\simeq -1+q(\tau) (2+\cos{\gamma}), 
  \label{mod-RG} \\
&&h'=-\frac{3}{2}h^2, \label{hubble}
\end{eqnarray}
where we define the time dependent parameter
\begin{equation}
  q(\tau)=\frac{1}{4}\lambda h^2.
  \label{q_eff}
\end{equation}
Since (\ref{hubble}) yields $h\propto 1/\tau$, 
$q(\tau)$ is the decreasing function of time. 
Defining the new variables $u,~v$
by $u=\Theta \cos{(\gamma/2)},~v=\Theta \sin{(\gamma/2)}$,  
(\ref{mod-RG}) is rewritten as follows:
\begin{equation}
\left(
\begin{array}{c}
u' \\
v'
\end{array}
\right)=\frac{1}{2}
\left(
\begin{array}{cc}
  0 & 1-q \\
3q-1 & 0
\end{array}
\right)
\left(
\begin{array}{c}
u \\
v
\end{array}
\right).
\label{matrix}
\end{equation}
The eigenvalue of the matrix in the right hand side of (\ref{matrix}) 
is real for $1/3\leq q\leq 1$ and $\Theta$ can grow during this epoch. 
We see that the function $q$ corresponds to the coefficient $q$ in the 
Mathieu equation (\ref{mathieu}). 
Recall that we have $\x\simeq h\cos{\tau},
~\y\simeq h\Theta\cos{(\tau+\gamma/2)}$ from (\ref{RG-sol}), 
we recognize that the amplitude of $\y$ 
has the growing behavior as the result of the 
parametric resonance. Soon after $q$ becomes smaller than $1/3$, 
$\Theta$ reaches a constant value. 
Fig.1 shows the evolution of the background fields obtained by 
numerical integration of (\ref{bg-eq1})(\ref{bg-eq2}).
The field $\y$ (the solid line) is amplified 
during the short period denoted by the horizontal arrow, 
which corresponds to the resonance epoch $1/3\leq q\leq 1$. 
We can evaluate how much $\y$ is amplified 
through the resonance epoch. 
From (\ref{mod-RG}), we have 
\begin{equation}
  \Theta\simeq \Theta_0 
  \exp{\left[\frac{1}{2}\int^{\tau}_{\tau_i} d\tau' q\sin{\gamma}\right]}.
  \label{zeta-reso}
\end{equation}
Using this expression, we get the upper bound
\begin{equation}
\frac{\Theta(\tau_f)}{\Theta(\tau_i)}\lesssim
\exp{(c_* \sqrt{\lambda})};~~
c_*=\frac{1}{6}\left(1-\frac{1}{\sqrt{3}}\right)\simeq 0.07, 
  \label{growth}
\end{equation}
where we used the relation $q(\tau_f)=1/3,~~q(\tau_i)=1$. 
The above expression is valid if the condition $\Theta\ll1$ 
is satisfied throughout the resonance epoch.
%
%
%
%%%%%%%%%%%%%%%%%%%%%%%%%%%%%%%%%%%%%%%%%%%%%%%%%%%%%%%%%%%%%%%%
\section{Evolution of the Bardeen parameter}
%%%%%%%%%%%%%%%%%%%%%%%%%%%%%%%%%%%%%%%%%%%%%%%%%%%%%%%%%%%%%%%%
%
%
%%%%%%% 
Behavior of the background solution obtained by RG method is translated 
to the metric perturbation $\zeta$ using the long wavelength 
solution (\ref{k=0-sol}). 
%%%%%%% 
The reduced system (\ref{mod-RG}) has the two integration constants: 
  The initial values of $\gamma,~\Theta$. During the resonance epoch, 
  $\gamma$ approaches a constant value which does not depend 
  on the initial value. We can regard $\Theta_0$ 
  (the initial value of $\Theta$) as the parameter $C$ described in 
  Sec.\ref{sec: k=0-sol}. That is, $\Theta_0$ is the parameter 
  which distinguishes the trajectories in the reduced system 
  $(\Theta,\gamma)$.
%%%%%%% 
Substituting (\ref{RG-sol}) into (\ref{k=0-sol}) and evaluating 
the integral adiabatically, we find that the dominant contribution 
to the Bardeen parameter $\zeta$ comes from the solution 
$Q_i^{ad}\equiv d\vpi/d\alpha$ and 
\begin{equation}
  Q_1^{iso}\equiv\frac{d\x}{dC}\simeq -h\Theta_{,C}
  \sin{\Theta}\cos{\tau},~~~
  Q_2^{iso}\equiv\frac{d\y}{dC}\simeq h\Theta_{,C} 
  \cos{\Theta}\cos{(\tau+\frac{\gamma}{2})}, 
  \label{iso-mode}
\end{equation}
where $\Theta_{,C}=\exp\left[\int d\tau'q/2\cdot \sin{\gamma}\right]$. 
The other independent solutions are recognized as the decaying modes
\cite{ataruya}. 

%%%%%%%
The solution $Q_i^{ad}$ is referred to as the adiabatic growing 
mode\cite{polarski}. 
Substituting $Q_i^{ad}$ into (\ref{bardeen2}), the Bardeen parameter 
remains constant in time. The constancy of $\zeta$ arises 
in the hydrodynamical perturbation without the entropy fluctuation
\cite{MFB}.
For the solution $Q_i^{iso}$, (\ref{bardeen2}) gives 
\begin{equation}
  \zeta^{iso}\simeq h\Theta_{,C} \sin{\Theta}\cos{\Theta}\cdot
\frac{\sin{\tau}\cos{\tau}-\sin{(\tau+\frac{\gamma}{2})}
\cos{(\tau+\frac{\gamma}{2})}}
{\cos^2{\Theta}\sin^2{\tau}+\sin^2{\Theta}\sin^2{(\tau+\frac{\gamma}{2})}}.
\label{zeta-iso}
\end{equation}
%
%%%%%%% 
We see that $\zeta^{iso}$ has the periodically sharp peaks around 
the zero points of $\x'$. This behavior also appears in the 
single field case\cite{KandH}. 
Except for the short interval of the peaks, 
$\zeta^{iso}$ traces the amplitude of $Q_1^{iso}$. 
Therefore $\zeta^{iso}$ deviates from zero 
when $\Theta_{,C}$ grows due to the parametric resonance. 
Note that $\zeta^{iso}$ vanishes when the background field $\y$ is 
zero ($\Theta=0$). 
Because the initial amplitude of the curvature perturbation $\zeta^{iso}$ 
is vanishingly small, we can identify $Q^{iso}_i$ is the 
contribution of the iso-curvature mode of the perturbation.

%%%%%%% 
We confirm the behavior of $\zeta^{iso}$ by solving 
(\ref{bg-eq1})(\ref{bg-eq2})(\ref{Mukhanov-eqs}) numerically. 
In Fig.2, we show the time evolution of the iso-curvature mode. 
In contrast to the constancy of the adiabatic growing mode 
(the dashed line), 
we observe that 
$\zeta^{iso}$ (the solid line) is amplified 
during the resonance epoch when the amplitude of the background 
$\y$ field grows (see Fig.1). After the resonance epoch, $\zeta^{iso}$ 
approaches zero  due to the Hubble damping which can be seen by 
the analytic result (\ref{zeta-iso}). 
%%%%%%% 
Using (\ref{growth})(\ref{zeta-iso}), we can estimate 
the growth of $\zeta^{iso}$ except for the sharp peak:
\begin{equation}
  \left|\frac{\zeta^{iso}(\tau_f)}{\zeta^{iso}(\tau_i)}\right|
  \lesssim  \frac{h(\tau_f)}{h(\tau_i)} \cdot \exp{(2 c_* \sqrt{\lambda})}.
  \label{zeta-max}
\end{equation}
% 
%
%
%
%
%
%%%%%%%%%%%%%%%%%%%%%%%%%%%%%%%%%%%%%%%%%%%%%%%%%%%%%%%%%%%%%%%%
\section{Summary and discussions}
%%%%%%%%%%%%%%%%%%%%%%%%%%%%%%%%%%%%%%%%%%%%%%%%%%%%%%%%%%%%%%%%
%
%
%%%%%%%%%
In this letter, we have investigated the cosmological perturbation 
of the model with two scalar fields in reheating phase after inflation.   
%%%%%%%%%
Using the background quantities, long wavelength solutions of 
the perturbation are obtained. Applying the RG method to the background 
dynamics in the coherent oscillating stage, we found that 
the massless field gets the effect of the parametric resonance. 
This leads to the amplification of the iso-curvature metric perturbations. 
The result implies that the Bardeen parameter deviates 
from the initial amplitude in reheating phase 
even though the initial amplitude determined in the inflationary stage  
is dominated by the adiabatic mode. 
Therefore the analysis using the scalar field model is essential 
to study the metric perturbations when the non-linear dynamics 
of the scalar fields plays an important role. 
%%%%%%%%%

The most important observed quantity is the power spectrum 
${\cal P}_{\zeta}(k)$. 
Consider the evolution of the metric fluctuation $\zeta(k)$ whose 
initial amplitude is given by the quantum fluctuation of the scalar 
fields during inflation.
As long as the wavelength of the fluctuation is outside the Hubble horizon, 
time evolution of $\zeta(k)$ is well-approximated by the solution for 
the long wavelength mode. 
The power spectrum for the present length-scale can be evaluated from 
the amplitude of $\zeta(k)$ when the fluctuations re-enter 
the Hubble horizon. 
Soon after the metric fluctuation $\zeta(k)$ suffers the 
parametric amplification during reheating, it approaches to the 
initial amplitude due to the Hubble damping.
The horizon re-entry time becomes earlier as the wavelength of the fluctuation 
becomes shorter. Therefore, we can expect that the effect of parametric resonance 
on the power spectrum at the horizon re-entry appears at the small scale (large $k$). 
We will give a detail discussion 
about evolution of the power spectrum in the forthcoming paper
\cite{ataruya}. 
%%%%%%%%%%%%%%%
%
%
%
\section*{ACKNOWLEDGMENT}
This work is partly supported by the Grand-In-Aid for 
Scientific Research of the Ministry of Education, Science, 
Sports and Culture of Japan $(09740196)$.
%%%%%%%%%%%%%%%%%%%%%%%%%%%%%%%%%%%%%%%%%%%%%%%%%%%%%%%%%%%%%%%%%%%%%%%%%%%%%%
%
%
%
%%%%%%%%%%%%%%%%%%%%%%%%%%%%%%%%%%%%%%%%%%%%%%%%%%%%%%%%%%%%%%%%
                        \appendix
%%%%%%%%%%%%%%%%%%%%%%%%%%%%%%%%%%%%%%%%%%%%%%%%%%%%%%%%%%%%%%%%
%
%
%
%%%%%%%%%%%%%%%%%%%%%%%%%%%%%%%%%%%%%%%%%%%%%%%%%%%%%%%%%%%%%%%%
\section*{RG method applying to the background system}
\label{app: RG}
%%%%%%%%%%%%%%%%%%%%%%%%%%%%%%%%%%%%%%%%%%%%%%%%%%%%%%%%%%%%%%%%
%
%
In this appendix, we apply the RG method to the background system 
(\ref{bg-eq1})(\ref{bg-eq2}). Following the treatment described 
in Sec.\ref{sec: bg-evolution}, we substitute (\ref{e-expansion}) 
to (\ref{bg-eq1})(\ref{bg-eq2}) and expand in powers of $\e$: 
\begin{eqnarray}
{\cal O}(\e^1)&:~~& \vx_{i(1)}^{\prime\prime}+\vx_{i(1)}=0,~~(i=1,2) 
\nonumber 
\\
&&h_{(1)}=\sqrt{\vx_{1(1)}^{\prime 2}+ 
\vx_{1(1)}^2+\vx_{2(1)}^{\prime 2}+\vx_{2(1)}^2},
\nonumber 
\\
{\cal O}(\e^2)&:~~& \vx_{1(2)}''+ \vx_{1(2)}=-3h_{(1)}\vx_{1(1)}'
-\tilde{\lambda} \vx_{2(1)}^2 \vx_{1(1)}, 
\nonumber 
\\
&& \vx_{2(2)}''+ \vx_{2(2)}=-3h_{(1)}\vx_{2(1)}'-\sigma \vx_{2(1)}
-\tilde{\lambda} \vx_{1(1)}^2 \vx_{2(1)}. 
\nonumber 
\end{eqnarray}
We first solve the above equations order by order. The solutions up to 
${\cal O}(\e^2)$ become
\begin{eqnarray}
\x&=&\e\left[A_0+\e(\tau-\tau_0)\left\{-\frac{3}{2}h_{(1)}A_0+
i\frac{\tilde{\lambda}}{2}\left(2|B_0|^2A_0 
+B_0^2A_0^*\right) \right\} \right]e^{i\tau}
\nonumber \\
&& 
+\e^2\frac{\tilde{\lambda}}{8}B_0^2A_0e^{i3\tau}+c.c.+{\cal O}(\e^3)
\nonumber 
\end{eqnarray}
and
\begin{eqnarray}
\y&=&\e\left[B_0+\e(\tau-\tau_0)\left\{-\frac{3}{2}h_{(1)}B_0
+i\frac{\tilde{\lambda}}{2}\left(2|A_0|^2B_0 
+A_0^2B_0^*\right)
+\frac{i}{2}\sigma B_0 \right\} \right]e^{i\tau}
\nonumber \\
&& +\e^2\frac{\tilde{\lambda}}{8}A_0^2B_0e^{i3\tau}+c.c.+{\cal O}(\e^3), 
\nonumber 
\end{eqnarray}
where $c.c.$ means complex conjugate and $h_{(1)}$ is given by 
\begin{equation}
  h_{(1)}=2\sqrt{|A_0|^2+|B_0|^2}.
\label{hy}
\end{equation}
$A_0,~B_0$ are the amplitudes at the initial time $\tau=\tau_0$.  
The above solutions are correct within $\e(\tau-\tau_0)\lesssim 1$ because 
the secular terms which grow as $(\tau-\tau_0)$ makes the perturbative 
expansion break down. To improve the perturbation series, we introduce 
an arbitrary parameter $\mu$. Splitting $\tau-\tau_0$ as 
$\tau-\mu+\mu-\tau_0$ and absorbing the terms containing $\mu-\tau_0$ into 
the renormalization constant $Z_A,~Z_B$ defined by 
$A_0=A(\mu)Z_A(\tau_0,\mu)$, $B_0=B(\mu)Z_B(\tau_0,\mu)$. This procedure is 
always possible for each order of expansion and we have the $\mu$ 
dependent solutions:  
\begin{eqnarray}
\x&=&\e\left[A+\e(\tau-\mu)\left\{-\frac{3}{2}h_{(1)} A+
i\frac{\tilde{\lambda}}{2}\left(2|B|^2A
+B^2A^*\right)\right\} \right]e^{i\tau}
\nonumber \\
&&
+\e^2\frac{\tilde{\lambda}}{8}B^2Ae^{i3\tau}+c.c.+{\cal O}(\e^3), 
\label{inf}
\\
\y&=&\e\left[B+\e(\tau-\mu)\left\{-\frac{3}{2}h_{(1)}B+
i\frac{\tilde{\lambda}}{2}\left(2|A|^2B
+A^2B^*\right)+\frac{i}{2}\sigma B\right\} \right]e^{i\tau}
\nonumber \\
&&
+\e^2\frac{\tilde{\lambda}}{8}A^2Be^{i3\tau}+c.c.+{\cal O}(\e^3). 
\label{chi}
\end{eqnarray}
Since $\mu$ does not appear in the 
original equations, the solution should not depend on $\mu$. This requires  
$(\partial\vpi/\partial \mu)_{\tau}=0$ for arbitrary $\tau$ and  
we obtain the RG equations: 
\begin{eqnarray}
  \frac{dA}{d\mu}&=&\e\left[-\frac{3}{2}h_{(1)} A+
i\frac{\tilde{\lambda}}{2}\left(2|B|^2A+B^2A^*\right)\right]+{\cal O}(\e^2), 
\nonumber \\
  \frac{dB}{d\mu}&=&\e\left[-\frac{3}{2}h_{(1)}B+\frac{i}{2}\sigma B
+i\frac{\tilde{\lambda}}{2}\left(2|A|^2B+A^2B^*\right)\right]
+{\cal O}(\e^2). 
\nonumber 
\end{eqnarray}
By setting $\mu\equiv\tau$ in (\ref{inf})(\ref{chi}), 
the secular terms are eliminated and we obtain the regular perturbation. 
Rewriting the variables $A,~B$ by 
\begin{eqnarray}
&&  A=\frac{1}{2}h_{(1)}\cos{\Theta}e^{i\theta},~~~
  B=\frac{1}{2}h_{(1)}\sin{\Theta}e^{i\psi}, 
\nonumber
\end{eqnarray}
and using (\ref{hy}), we get the solution (\ref{RG-sol}) and 
the above RG equations reduce to
\begin{eqnarray}
&&\Theta'= \e
\frac{\tilde{\lambda}}{16}h_{(1)}^2 \sin{\gamma}\sin{2\Theta},
  \label{RG-1} \\
&&\gamma'= -1 + \e\frac{\tilde{\lambda}}{4}h_{(1)}^2 
\cos{2\Theta}(2+\cos{\gamma}),
  \label{RG-2} \\
&&\theta'= \e\frac{\tilde{\lambda}}{8}h_{(1)}^2 
\cos{\gamma}\sin^2{\Theta},~~~~~
\gamma=2(\psi-\theta), 
  \label{RG-3} \\
&&h_{(1)}'=-\e\frac{3}{2}h_{(1)}^2, 
  \label{RG-4} 
\end{eqnarray}
where we set $\e\sigma=-1$.
%
%
%%%%%%%%%%%%%%%%%%%%%%%%%%%%%%%%%%%%%%%%%%%%%%%%%%%%%%%%%%%%%%%%%%%%%%%%%%%%
%%%%%%%%%%%%%%%%%%%%%%%%%%%%%%%REFERENCES%%%%%%%%%%%%%%%%%%%%%%%%%%%%%%%%%%%
%%%%%%%%%%%%%%%%%%%%%%%%%%%%%%%%%%%%%%%%%%%%%%%%%%%%%%%%%%%%%%%%%%%%%%%%%%%%
%
%
%
%
\baselineskip23pt
%
%

%
%
%
%%%%%%%%%%%%%%%%%%%%%%%%%%%%%%%%%%%%%%%%%%%%%%%%%%%%%%%%%%%%%%%%%%%%%%%%%%%
%
%
%
%
\newpage
%
%%%%%%%%%%%%%%%%%%%%%%%%%%%%%%%%%%%%%%%%%%%%%%%%%%%%%%%%%%%%%%%%%%%%%%%%%%%
%%%%%%%%%%%%%%%%%%%%%%%%%%%%%%FIGURE CAPTION%%%%%%%%%%%%%%%%%%%%%%%%%%%%%%%
%%%%%%%%%%%%%%%%%%%%%%%%%%%%%%%%%%%%%%%%%%%%%%%%%%%%%%%%%%%%%%%%%%%%%%%%%%%
%
%
\begin{center}
  {\bf FIGURE CAPTIONS \\}
\end{center}
\begin{description}
%%%%%%%%%%%%%%%%%%%%%%%%%%%%%%%%%%%%%%%%%%%%%%%%%%%%%%%%%%%%%%%%%%%%%%%%%%%%
\item[Fig.1] The amplitude of the 
background fields $\x$ and $\y$ is shown as a function of time $\tau$ by 
solving (\ref{bg-eq1})(\ref{bg-eq2}) numerically. The broken and 
the solid lines represent the inflaton($\x$) and the massless field($\y$) 
respectively. During the period denoted by the horizontal 
arrow, the field $\y$ is amplified by the effect of parametric resonance. 
We used the initial values which correspond to the expression 
(\ref{RG-sol}) at $\tau=0$ with the parameters 
$\lambda=4000,~h=0.035,~\Theta=0.01,~\gamma=0.4,~\theta=0.0$. 
%%%%%%%%%%%%%%%%%%%%%%%%%%%%%%%%%%%%%%%%%%%%%%%%%%%%%%%%%
\item[Fig.2] The Bardeen parameter $\zeta^{iso}$ is 
plotted as a function of time $\tau$ by solving 
(\ref{bg-eq1})(\ref{bg-eq2})(\ref{Mukhanov-eqs}) numerically. 
We calculate (\ref{bg-eq1})(\ref{bg-eq2}) 
with the same initial condition as described in Fig.1. As for 
the perturbations, we set the initial values corresponding 
to the expression (\ref{iso-mode}) at $\tau=0$. 
The figure shows that $\zeta^{iso}$ (the solid line) is amplified 
during the resonance epoch denoted by the horizontal arrow 
and approaches zero due to the Hubble damping. 
We also plotted the Bardeen parameter for the 
adiabatic growing mode (the dashed line) obtained by the substitution of 
$Q^{ad}_i$ into (\ref{bardeen2}).

\end{description}
%
%
%%%%%%%%%%%%%%%%%%%%%%%%%%%%%%%%%%%%%%%%%%%%%%%%%%%%%%%%%%%%%%%%%%%%%%%%%%%
%
%
%
%
%%%%%%%%%%%%%%%%%%%%%%%%%%%%%%%%%%%%%%%%%%%%%%%%%%%%%%%%%%%%%%%%%%%%%%%%%%%
%%%%%%%%%%%%%%%%%%%%%%%%%%%%%%%%FIGURES%%%%%%%%%%%%%%%%%%%%%%%%%%%%%%%%%%%%
%%%%%%%%%%%%%%%%%%%%%%%%%%%%%%%%%%%%%%%%%%%%%%%%%%%%%%%%%%%%%%%%%%%%%%%%%%%
%
\newpage
\pagestyle{empty}

%\begin{figure}[p]
\epsfxsize=\hsize
\epsfbox{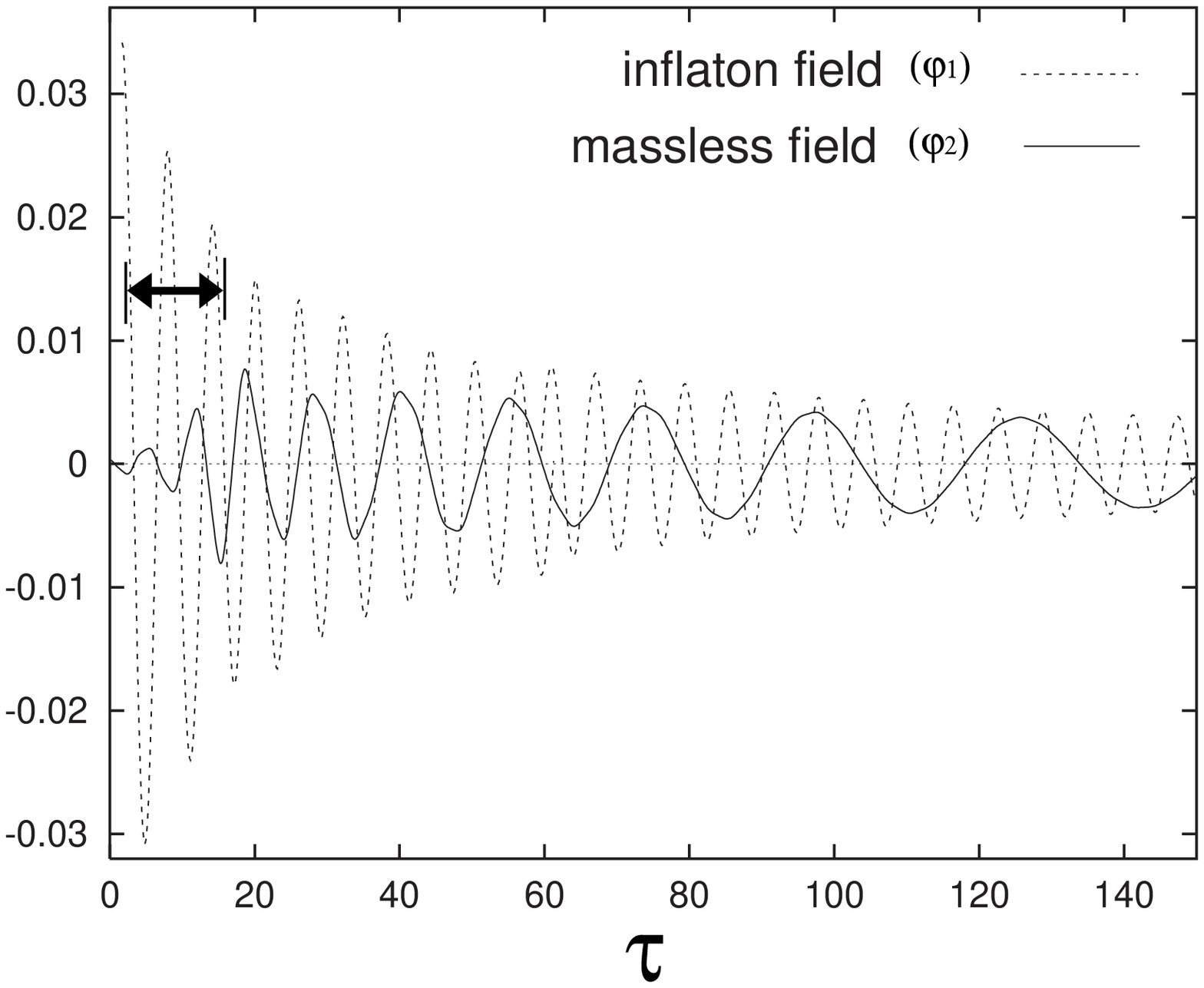}
\vspace{2.5cm}
\begin{center}
{\large Fig.1}
\end{center}
%\caption{}
%\end{figure}

%%%%%%%%%%%%%%%%%%%%%%%%%%%%%%%%%%%%%%%%%%%%%%%%%%%%%%%%%%%%%%%%%%%%%%%%%%%
%\begin{figure}[p]
\epsfxsize=\hsize
\epsfbox{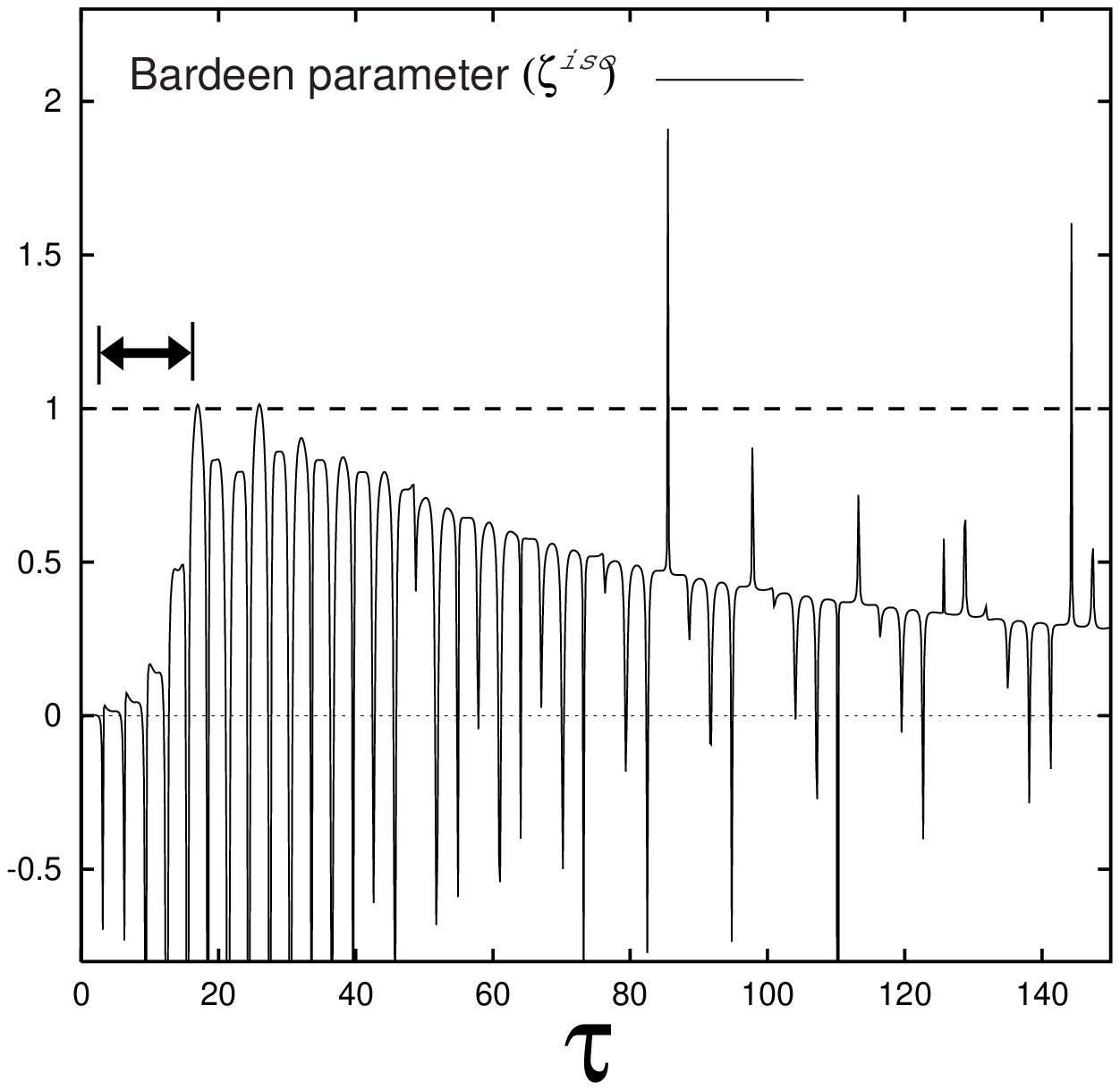}
\vspace{2.5cm}
\begin{center}
{\large Fig.2}
\end{center}
%\caption{}
%\end{figure}

%%%%%%%%%%%%%%%%%%%%%%%%%%%%%%%%%%%%%%%%%%%%%%%%%%%%%%%%%%%%%%%%%%%%%%%%%%%

\begin{references}
\bibitem{Dolgov} A.Dolgov and D.Kirilova, 
  Sov.J.Nucl.Phys.{\bf 51} (1990) 172.
\bibitem{Traschen} J.Traschen and R.Brandenberger, 
  Phys.Rev.D {\bf 42} (1990) 2491.
\bibitem{KFL1} L.Kofman, A.Linde and A.A.Starobinsky, 
  Phys.Rev.Lett.{\bf 73} (1994) 3195.
\bibitem{STB} Y.Shtanov, J.Traschen and R.Brandenberger, 
  Phys.Rev.D{\bf 51} (1995) 5438.
\bibitem{KFL2} e.g, L.Kofman, A.Linde and A.A.Starobinsky, 
  Phys.Rev.D {\bf 56} (1997) 3258 and references there in.
\bibitem{ynambu}  Y.Nambu and A.Taruya, Prog.Theor.Phys.{\bf97} (1997) 83. 
\bibitem{KandH} H.Kodama and T.Hamazaki, Prog.Theor.Phys.{\bf96} (1996) 949.
\bibitem{hwang1}  J.Hwang, Phys.Lett.{\bf B401} (1997) 241.
\bibitem{HandK} T.Hamazaki and H.Kodama, Prog.Theor.Phys.{\bf96} (1996) 1123.
\bibitem{Lyth} D.Lyth, Phys.Rev.D {\bf 31} (1985) 1792.
\bibitem{Salopek} D.S.Salopek, J.R.Bond and J.M.Bardeen, 
  Phys.Rev.D {\bf 40} (1989) 1753.
\bibitem{MFB} V.F.Mukhanov, H.A.Feldman and R.H.Brandenberger, 
  Phys.Rep. {\bf 215} (1992) 203.
\bibitem{BandW}   J.G.Bellido and D.Wands, Phys.Rev.D {\bf 53} (1996) 5437.
\bibitem{Mukhanov} V.F.Mukhanov, Sov.Phys.JETP {\bf 67} (1988) 1297.
\bibitem{hwang2} J.Hwang, preprint gr-qc/ 9608018.

\bibitem{Oono} L.Y.Chen, N.Goldenfeld and Y.Oono, Phys.Rev.E {\bf 54} (1996) 
  376.
\bibitem{kunihiro}  T.Kunihiro, Prog. Theor.Phys.{\bf 94} (1995) 503. 
\bibitem{polarski} D.Polarski and A.A.Starobinsky, Nucl.Phys.{\bf B385} 
(1992) 623; Phys.Rev.D {\bf 50} (1994) 6123.
\bibitem{ataruya} A.Taruya and Y.Nambu, in preparation.
\end{references}
\end{document}